\documentstyle[12pt,draft,aas_macros]{nature}
\font\large=cmbx10 scaled \magstep3

\newcommand{\be}{\begin{equation}}
\newcommand{\ee}{\end{equation}}

\newcommand{\ifm}[1]{\relax\ifmmode#1\else$\mathsurround=0pt #1$\fi}
\newcommand{\kms}{\ifmmode\,{\rm km}\,{\rm s}^{-1}\else km$\,$s$^{-1}$\fi}
\newcommand{\kpc}{\ifmmode\,{\rm kpc}\else kpc\fi}
\newcommand{\Mpc}{\ifmmode\,{\rm Mpc}\else kpc\fi}

\newcommand{\ltsima}{$\; \buildrel < \over \sim \;$}
\newcommand{\lsim}{\lower.5ex\hbox{\ltsima}}
\newcommand{\gtsima}{$\; \buildrel > \over \sim \;$}
\newcommand{\gsim}{\lower.5ex\hbox{\gtsima}}

\begin{document}

\title{Resonant stripping\\ 
as the origin of dwarf spheroidal galaxies}

%% Notice placement of commas and superscripts and use of &
%% in the author list

\author{Elena D'Onghia$^{1,2}$, 
        Gurtina Besla$^{2}$, 
        Thomas J. Cox$^{2}$
        \& Lars Hernquist$^2$
\institute{$^1$Institute for Theoretical Physics,
              University of Zurich, CH-8057 Zurich, Switzerland. \\
$^2$Harvard-Smithsonian Center for Astrophysics, 60 Garden St. MS
   51, Cambridge, MA, 02138.\\
}
}
\date{\today}{}
\headertitle{Resonant Stripping}
\mainauthor{D'Onghia et al.}

\summary{
Dwarf spheroidal galaxies are the most dark matter dominated systems
in the nearby Universe\cite{Kleyna,Chapman,Wilkinson} 
and their origin is one of
the outstanding puzzles of how galaxies form.  Dwarf spheroidals are
poor in gas and stars, making them unusually 
faint\cite{Mateo,Grebel,Gallagher},
and those known as ultra-faint dwarfs\cite{Willman,Zucker} have by far the
lowest measured stellar content of any galaxy\cite{Penarrubia,Strigari}.  
Previous
theories\cite{Mayer} require that dwarf spheroidals orbit near giant
galaxies like the Milky Way, but some dwarfs have been observed in the
outskirts of
 the Local Group\cite{Grebel03}.  Here we report simulations
of encounters between dwarf disk galaxies and somewhat larger objects. 
We find that the encounters excite a process, which we term ``resonant stripping'', that 
can transform them into dwarf spheroidals.  This effect is
distinct from other mechanisms proposed to form dwarf spheroidals,
including mergers\cite{Toomre77}, galaxy-galaxy harassment\cite{Moore96},
or tidal and ram pressure stripping, because it is driven by gravitational resonances.  
It may account for the observed properties of dwarf spheroidals in the Local Group, including their
morphologies and kinematics.
Resonant stripping predicts that dwarf
spheroidals should form through encounters, 
leaving detectable long stellar streams and tails.}

\maketitle

In the widely accepted cold dark matter theory, bound systems form
hierarchically through gravitational collapse.  Because gravity acts
in the same manner on all forms of mass, it is expected that galaxies
should contain a ratio of dark to ordinary luminous material roughly
similar to the cosmic mean.  However, dwarf spheroidals have
much larger dark to luminous mass ratios\cite{Kleyna,Chapman,Wilkinson}, 
requiring
some mechanism to efficiently separate gas and stars from dark matter
in order to explain their origin.  Two classes of models have been
proposed to account for the properties of dwarf spheroidals.  In one class,
gas is either prevented from collecting in shallow 
potential wells of dark matter by photoheating during cosmic reionization or blown
out subsequently by feedback from star formation before a significant
mass in stars can accumulate\cite{Bullock,Dekel}.  However, there are no
clear signatures of reionization in the star formation histories of
these galaxies\cite{Grebel03}, which in many cases continued to form stars
beyond the reionization epoch, and for most of them the dark halo
masses are likely too large (10$^8$-10$^9$
M$_{\odot}$\cite{Wilkinson,Penarrubia,Strigari}) 
for supernova-driven winds to remove
gas\cite{MacLow}.

\begin{figure}
\vskip 10.0cm
{\includegraphics{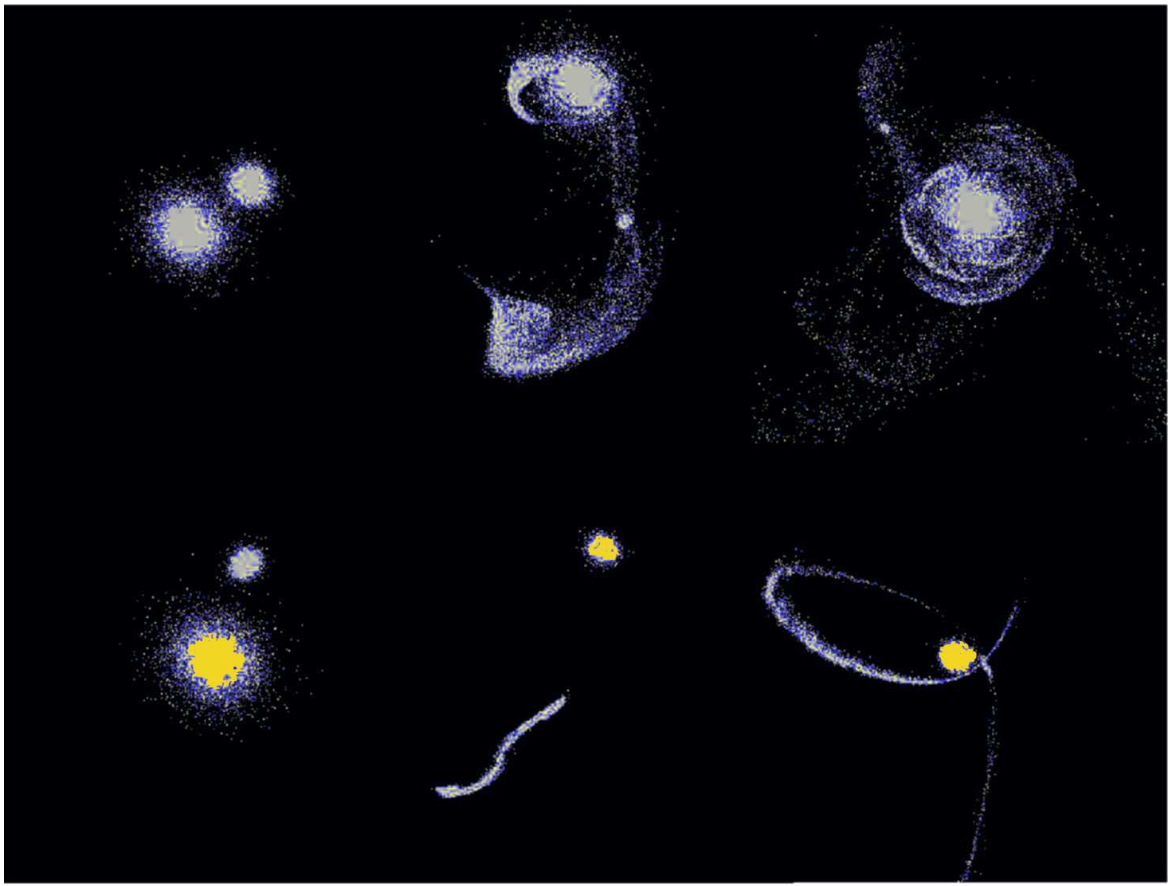}}
\vskip 0.2cm
\caption[]{\small
{\bf Encounters between galaxies.}
{\it Top Row:} Interaction between a dwarf galaxy with a mass of
 1.7x10$^8$ M$_{\odot}$ orbiting around a larger dwarf with 100 times its mass. 
Only the stellar components are plotted. 
The upper left panel illustrates the initial set up where the two dwarfs
approach one another on a somewhat prograde orbit (the disks are seen face on).  The upper
middle panel gives the state of the system after 2 billion years, following the
first pericentric passage, and the upper right panel shows the appearance of
the galaxies after 7 billion years.  Each panel displays a region 100
kpc on a side. An outcome like the one illustrated in the top panels of Figure 1
occurs preferentially when one of the interacting galaxies is between 10 and 100
times more massive than the other one.
If the galaxies have nearly the same mass they will merge quickly,  
masking the effects of resonant stripping because nearly all the 
luminous matter will remain bound to the remnant.
{\it Bottom Row:} Shown is the orbit of the same small galaxy (in white)  
around the Milky Way today (in yellow), which has 10,000 times its mass.
Although the encounter is mostly prograde, the spin and
orbital frequencies are no longer well-matched and the resonant 
response is suppressed. The left panel displays a region 150
kpc on a side and the initial setup;  the middle and right panels show an expanded view 300 kpc on a side
and give the state of the system after 2 and 7 billion years, respectively.
}
\end{figure}

The other class of models relies on impulsive heating by gravitational tidal
forces to strip stars and reshape the stellar component of a
disk-dominated dwarf galaxy.  Variously referred to as ``tidal
stirring''\cite{Mayer} or ``tidal shocking''\cite{Gnedin}, this process
operates when the gravitational potential acting on a small galaxy
varies rapidly as it orbits in a larger system and can, in principle,
convert a rotationally supported disk of stars into a spheroid.
However, gas can radiatively dissipate energy; thus, scenarios based
on gravitational heating require an additional process, i.e. ram
pressure stripping, to remove the gas\cite{Mayer}.  Because
gravitational heating and ram pressure stripping occur with varying
efficiencies, models coupling them may rely on particular conditions
to produce dwarf spheroidal galaxies.

This second class of models also requires that dwarfs orbit close to
giant galaxies.  Whereas several of the brightest dwarf spheroidals in
the Local Group cluster around the large galaxies, some, like Cetus
and Tucana, reside at great distances from both the Milky Way and
Andromeda\cite{Gallagher,Sales}.  
Furthermore, many nearby dwarf spheroidals lie
in the same plane as the orbits of the Magellanic Clouds and the
Magellanic Stream\cite{Lynden-Bell}. A thin planar distribution may be
difficult to explain if dwarf galaxies fall in
individually\cite{Libeskind,Zentner} but 
can arise naturally if dwarfs fall into
larger systems as members of groups of dwarfs\cite{D'Onghia}.  Examples of
such groups have been found nearby\cite{Tully}. Moreover,
interactions between dwarfs should be common in these environments,
especially in the past when the Universe was younger and more dense.

\begin{figure}
\vskip 12.5cm
{\includegraphics{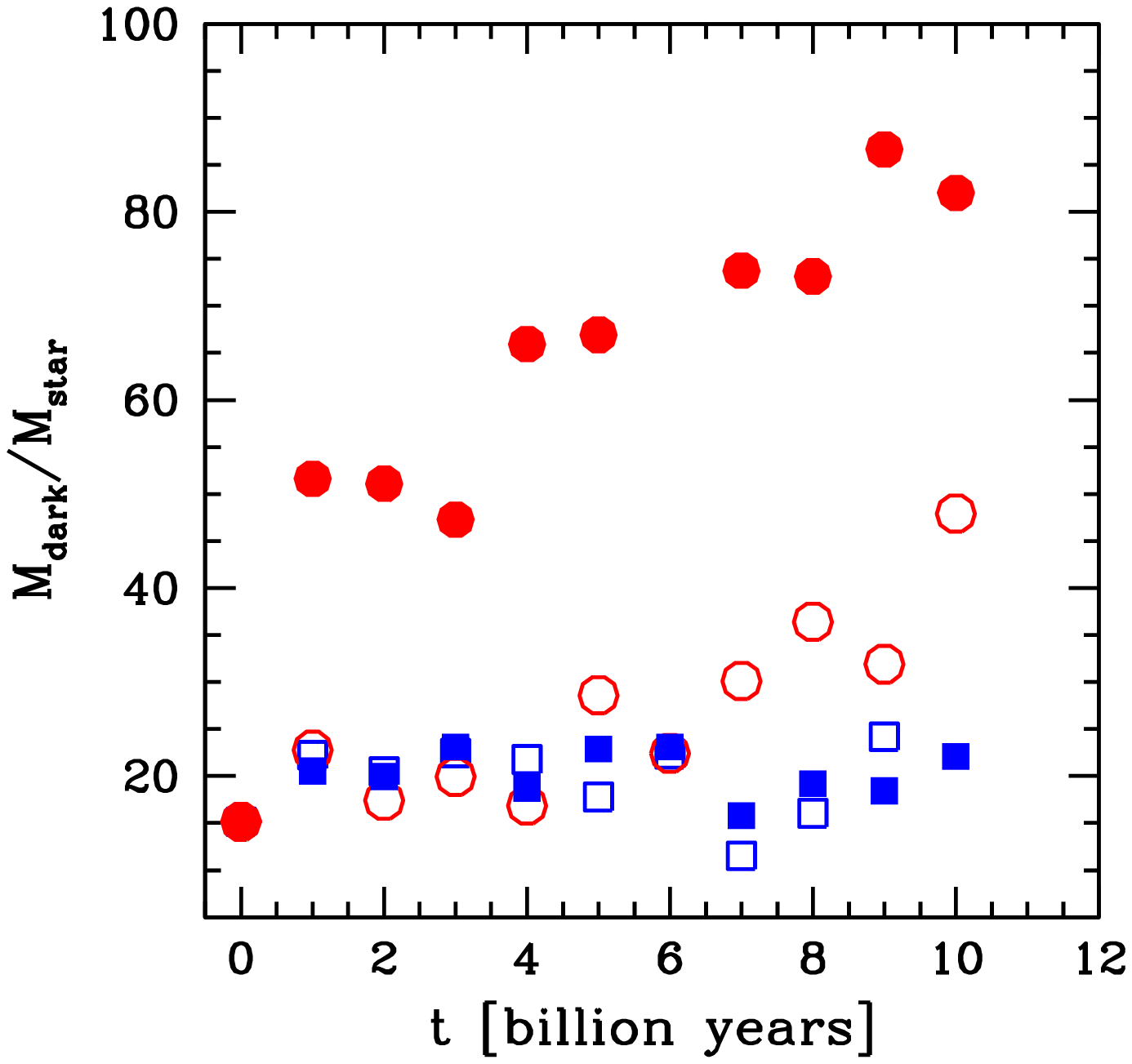}}
\vskip -6cm
\caption[]{\small 
{\bf The time evolution of the dark-to-luminous mass ratio 
(M$_{\rm{dm}}$/M$_{\rm{star}}$)}.
The  M$_{\rm{dm}}$/M$_{\rm{star}}$
of the smaller dwarf being resonantly stripped is computed at the tidal radius and marked by the filled red circles. 
The same case but for a mostly 
retrograde encounter is 
illustrated by the open red circles. A resonance occurs if
$\Omega_s = \Omega_0 \ \ \rightarrow \ \ \frac{v}{r} \sim \frac{V_0}{R_{\rm{peri}}}\sqrt{(1+e)},$
where $v$ is the rotation velocity, 
$r$ is the size of the smaller dwarf, 
$V_0$ is the orbital velocity, $R_{\rm{peri}}$ the pericentric distance, 
and $e$ the eccentricity of the orbit. Note that this resonance condition
is not dependent on the specific choice of the pericentric 
distance alone but rather on the {\it combination} of the internal structure 
(e.g. the rotation curve) of the small dwarf and the orbital parameters.
In other words, if the pericentric distance changes, the resonance condition could  
still be satisfied provided that the disk rotation speed were modified accordingly.
If the orbit is more retrograde, stars are not preferentially removed immediately. So,  
after 2 billion years, the net change in M$_{\rm{dm}}$/M$_{\rm{star}}$ is a factor of
4 larger for the prograde versus retrograde cases illustrated. 
However, after 4 billion years the internal structure of the smaller
dwarf and the orbit are affected by
gravitational torques, allowing resonant stripping to occur.
The ratio M$_{\rm{dm}}$/M$_{\rm{star}}$ of the small galaxy orbiting about the Milky Way 
today is plotted
for prograde (filled blue squares) and retrograde (open blue squares) orbits.
In both these cases, the spin and orbital frequencies of the galaxies are no
longer comparable and the resonant interaction is
suppressed, even in the prograde case. }
\end{figure}

Here we describe numerical experiments to investigate the consequences
of encounters between dwarf galaxies.  Figure 1 shows
the time evolution of an interaction between a pair of dwarf galaxies.
This could represent an interaction between dwarfs in a small group or
between a dwarf and the forming Milky Way at high redshift.
After 2 billion years nearly $\sim$80\% of the stars are stripped away
from the smaller dwarf but its surrounding dark matter halo is less
strongly affected, leading to a change in the ratio of dark to
luminous matter.  This outcome is surprising because the stars sit at
the bottom of the potential well and therefore comprise the most
tightly bound material in the galaxy.

This unexpected result is caused by a gravitational process, which we
term ``resonant stripping''.  The efficient removal of stars from the
smaller dwarf over the course of the interaction shown in Figure 1 is
mediated by a resonance between the spin angular frequency
($\Omega_s$) of its disk and the angular frequency of its orbit
($\Omega_0$) around the larger dwarf.  When these frequencies are
comparable in magnitude and the spin and orbital angular momenta are
somewhat aligned (i.e. a prograde encounter), the gravitational
perturbation from the larger dwarf on a patch of the smaller dwarf's
disk is always directed outwards from its center, leading to
significant stripping.  However, for a perfectly retrograde collision,
stars in the disk are alternately pulled inward and outward with
little net result.  The minimal response from the dark matter is
expected because these particles move on random orbits, and the net
perturbation on the halo mostly averages out.

\begin{figure}
\vskip 12.5cm
{\includegraphics{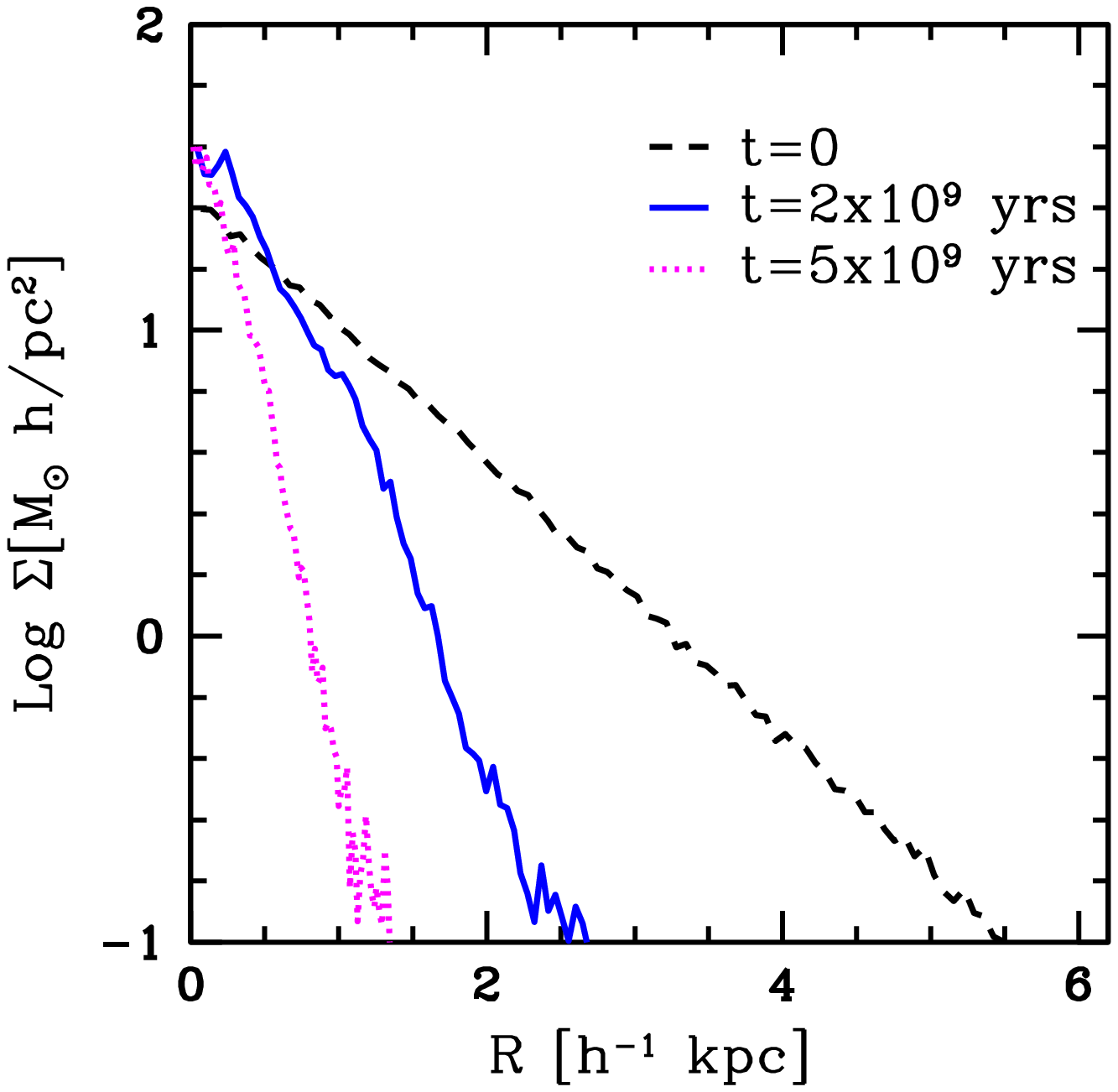}}
\vskip -5cm
\caption[]{\small 
{\bf Radial stellar surface mass density profile 
of the smaller dwarf.} The profile is 
plotted for a prograde encounter 
at the initial time of the
simulation (dashed black line), after the first pericentric passage (2 billion years;
solid blue line), and after 5 billion years (magenta dotted line). 
The profile is initially an exponential distribution with effective radius, R$_e$,
appropriate for dwarf disk
galaxies. However, it evolves immediately into a more concentrated  
profile with R$_{e}\sim 0.5 \ h^{-1}$ kpc (blue line) 
and after 5 billion years the disk is converted into a compact spheroid with
a smaller effective radius.}
\end{figure}

\begin{figure}
\vskip 12.5cm
{\includegraphics{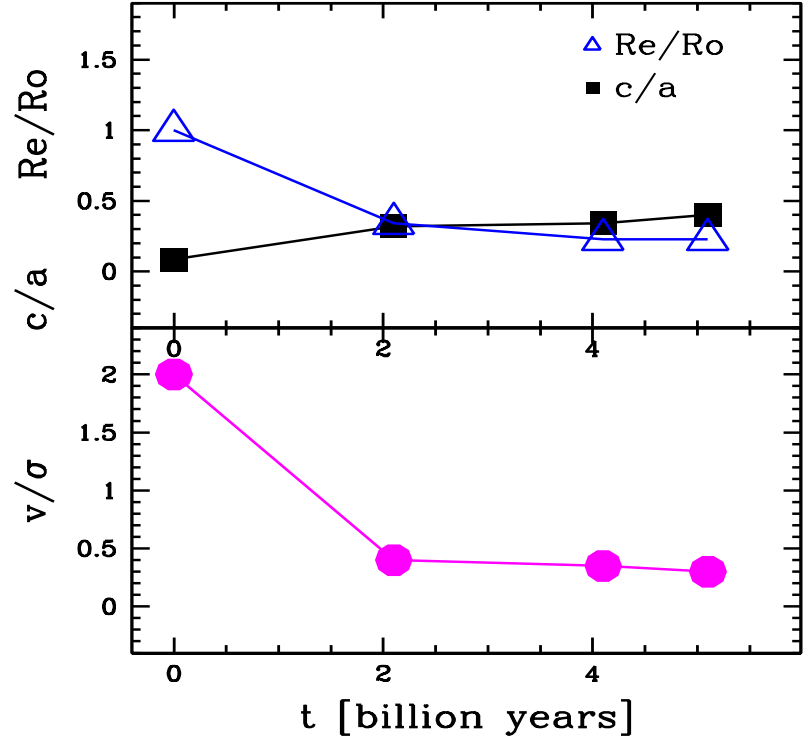}}
\vskip -5cm
\caption[]{\small 
{\bf The time evolution of the structural properties of the small 
dwarf galaxy.} The morphological evolution of the small 
dwarf galaxy over the course of the prograde encounter
is quantified at characteristic
times (0,2,4 and 5 billion years). The top panel illustrates the evolution of the
effective radius, R$_{e}$, normalized to its value at the beginning of the simulation
and is marked by blue open triangles. The evolution of the minor to
major axis ratio of the bound stellar component ($c/a$) measured at R$_{e}$ is
plotted as filled squares in the same panel. 
The squares indicate that the initial stellar disk
($c=0$) changes its shape immediately after 
2 billion years into a more spheroidal object ($c\approx 0.5$).
This evolution owes to the complete removal of the outer stellar disk via resonant
stripping.
The bottom panel shows the evolution of the kinematic properties
of the dwarf, measured in terms of the ratio of the
centrifugal rotational velocity of the small dwarf, $v$, to its random
motions, $\sigma$. The ratio $v/\sigma$ (closed circles) evolves after the
first pericentric passage (2 billion years) 
indicating a change  from a state where the rotational
support at R$_{e}$ is twice that from random motions to a value for $v/\sigma$
less than 0.5 - such a low value is similar to that of dwarf
spheroidals not only in the Local Group but also in galaxy clusters\cite{Ferguson}.}
\end{figure}

Although resonances have previously been invoked to produce bridges and
tails in collisions between galaxies of comparable
mass\cite{Toomre}, our work shows that resonant stripping can alter
the mass to light ratio of dwarfs by removing luminous material more
efficiently than dark matter. In Figure 2 the luminous fraction of the
smaller dwarf is plotted as a function of time during its interaction
with a galaxy 100 times its mass (filled circles), compared to another
simulation designed to follow a similar dwarf orbiting about the Milky
Way today (filled squares). Resonant stripping is most effective for
galaxies interacting on prograde orbits, as in the example shown in
the top panels in Figure 1.  
However, our calculations further
indicate that gravitational torques can alter the internal structure
of the galaxies and the relative alignments of spin and orbital
angular momenta, allowing the resonance to eventually act even for
orbits that are initially somewhat retrograde (see open circles
in Figure 2).

Resonant stripping is distinct from other processes proposed to drive
galaxy evolution, such as mergers\cite{Toomre77}, galaxy-galaxy
harassment\cite{Moore96} or more general heating processes, and tidal or
ram pressure stripping.  In particular, mechanisms that can be treated
using the impulse approximation do not account for resonances because
the particles in the perturbed system are assumed to remain roughly
stationary over the course of the encounter.  Because resonant
stripping will affect gas and stars in a similar manner in a
rotationally supported disk, it is simpler than models that require
separate effects to strip the gas versus the stars.

Resonant stripping can drive the morphological evolution of dwarfs.
When operating in low mass groups, this mechanism can pre-process
dwarfs by transforming disk galaxies into spheroids before they are
accreted by larger galaxies like the Milky Way.  We find that dwarf
spheroidal galaxies formed in this manner have properties similar to
those of dwarfs observed in the Local Group.  This is demonstrated in
Figures 3 and 4, which show, respectively, that the final radial
surface mass profiles and kinematic properties of a dwarf disk galaxy
undergoing resonant stripping in our simulations are similar to those
of observed dwarf spheroidals in the Local Group\cite{Ferguson}.

Our model makes definite predictions that can be tested in the future.
In particular, resonant stripping should be visible in situ in
associations of dwarfs.  Unlike in previous theories, dwarf
spheroidals are thus expected to be found along
with detectable stellar tails and shells, marking their formation.  If
this is indeed their dominant production mechanism, our model predicts
that dwarf spheroidals should have similar properties in different
environments, which is supported by the observed similarities between
dwarf spheroidals in the Perseus cluster\cite{Penny} and those in the
Local Group.

The efficiency of resonant stripping depends on various factors,
including the internal structure of the smaller dwarf (see Supplementary
Information).  In the example
shown in Figure 1, the interacting dwarfs were chosen to have
characteristics similar to those of local dwarf disk galaxies.
However, the detailed properties of young low-mass galaxies formed at
early cosmic times are unknown, here we are assuming 
they are cold disks.  If the smaller dwarf in our
simulations was set up so as to satisfy the condition for resonant
stripping even more precisely and over a larger portion of the disk, it
is possible that more luminous material could be
stripped, depleting the stellar surface density in the inner regions more
significantly than in the example shown in Figure 3. In that event,  
resonant stripping might  yield ultra-faint dwarf galaxies or even systems
that are nearly entirely dark.  This leads us to speculate that, when
placed in a proper cosmological framework, resonant stripping might
explain the missing satellite problem in the Local
Group\cite{Klypin,Moore}.  Moreover, spectroscopic measurements indicate
that the Virgo cluster is also missing satellite galaxies\cite{Rines}.
Resonant stripping acting in low mass groups when the Universe was
younger may have caused halos to lose their gas and stars,
pre-processing them before these groups were accreted into galaxy
clusters, providing a similar evolutionary mechanism to reproduce the
luminosity functions of both galaxy groups and clusters.

\vskip 0.5cm
\begin{acknowledge}
This research  was partly supported by the EU Marie Curie 
Intra-European Fellowship under contract MEIF-041569 and from NSERC
postgraduate fellowship.
Numerical simulations were performed on the Odyssey supercomputer
at Harvard University.
\end{acknowledge}

\vskip 0.5cm
\noindent{\bf Author Information} The authors declare that they have no competing financial
interests. Correspondence and requests for materials should be addressed
E.D. (edonghia@cfa.harvard.edu).
\newpage

%%
%% TABLES
%%
%% If there are any tables, put them here.
%%

\begin{center}
\vspace*{-10.00pt}
{\Large \bf SUPPLEMENTARY INFORMATION}
\end{center}

\noindent
This is an extension of the Letter to Nature,
aimed at providing further details, in support of the results reported 
in the main body of the Letter.  

\section{Numerical Methods}

The simulations were carried out with GADGET3, a parallel
TreePM-Smoothed particle hydrodynamics (SPH) code developed to compute
the evolution of stars and dark matter, which are treated as collisionless
fluids. The six-dimensional phase space is discretized into fluid
elements that are computationally realized as particles in the
simulations. A detailed description of the code is available in the
literature\cite{S05,Setal08}. Here we note its essential features.

GADGET3 is a cosmological code in which the gravitational field on
large scales is calculated with a particle-mesh (PM) algorithm, while
the short-range forces are computed using a tree-based hierarchical
multipole expansion, resulting in an accurate and fast gravitational
solver.  The scheme combines the high spatial resolution and relative
insensitivity to clustering of tree algorithms with the speed and
accuracy of the PM method to calculate the long range gravitational
field.

Pairwise particle interactions are softened with a spline
kernel\cite{HK89} of scalelength $h_s$, so that they are strictly
Newtonian for particles separated by more than $h_s$.  The resulting
force is roughly equivalent to traditional Plummer-softening with
scalelength $\epsilon \approx h_s /2.8$. For our applications the
gravitational softening length is fixed to $h_s$=70 $h^{-1}$ pc throughout
the evolution of the galaxy encounters. (Here, $h$ is the
Hubble constant in units of 100 km/s/Mpc.)

At each output time, we measure the total mass of the dark halo of
each galaxy using the SUBFIND algorithm\cite{Setal08}. This code
identifies as halo members particles in groups that are
gravitationally self-bound and are overdense with respect to the local
background.

\section{Setting the initial conditions}

Each galaxy in our study consists of a dark matter halo and a
rotationally supported disk of stars. The parameters describing each
component are independent and models are constructed in a manner
similar to the approach described in previous
works\cite{H93,S00,SDH04}.

\section{Dark Halo}
We model the dark matter mass distribution of each galaxy with a Hernquist\cite{H0} profile:
\begin{equation}
\rho_{\rm{dm}}=\frac{M_{\rm{dm}}}{2\pi}\frac{a}{r(r+a)^3}
\end{equation}
with cumulative mass 
distribution $M(<r)=M_{\rm{dm}}r^2/(r+a)^2$, where $a$
is the radial scale length.

We motivate this choice for the halo profile 
by the fact that in its inner parts
the shape of this profile is identical to the NFW\cite{NFW97} fitting formula 
for the mass density distribution of dark matter halos in cosmological
simulations. However, owing to its faster decline in the outer parts,
the total mass described by the Hernquist model converges, allowing
the construction of isolated halos without the need of an {\it ad hoc}
truncation. To make the connection with common descriptions
of halos in cosmological simulations we associate the Hernquist profile
for the galaxy halo with a corresponding NFW-halo profile such that the
inner density profiles are equivalent at radii less than the virial
radius, r$_{200}$,the radius
at which the mean enclosed dark matter density is 200 times the critical
density\cite{NFW97}.
This implies a relation between $a$ and the scale length $r_s$ 
of the NFW profile. The latter is often given
in terms of the concentration index $c$, conventionally defined 
as $c=r_{200}/r_s$.
Hence, we adopt the relation\cite{SDH04}:\\

\begin{equation}
a=r_s\sqrt{2[\rm{ln}(1+c)-c/(1+c)]} .
\end{equation}
For the small dwarf galaxy, we choose a concentration index of c=25,
which is a value motivated by cosmological simulations\cite{Neto07}
and a spin parameter of 0.09.
Its corresponding radial scalelength $a$ is 0.85 $h^{-1}$ kpc.
The tidal radius for the small dwarf galaxy computed at the  
initial distance from the larger dwarf, 30 $h^{-1}$ kpc,  
is 3  $h^{-1}$ kpc. This value is almost 30\% of the virial radius
and the dark matter mass of the little dwarf galaxy 
within this radius is $\sim 15$ times larger than the stellar 
mass disk. The tidal radius changes by time along the orbit 
and at the pericentric distance is $\sim$ 1.2 $h^{-1}$ kpc.

\begin{figure}
\vskip 8.5cm
{\includegraphics{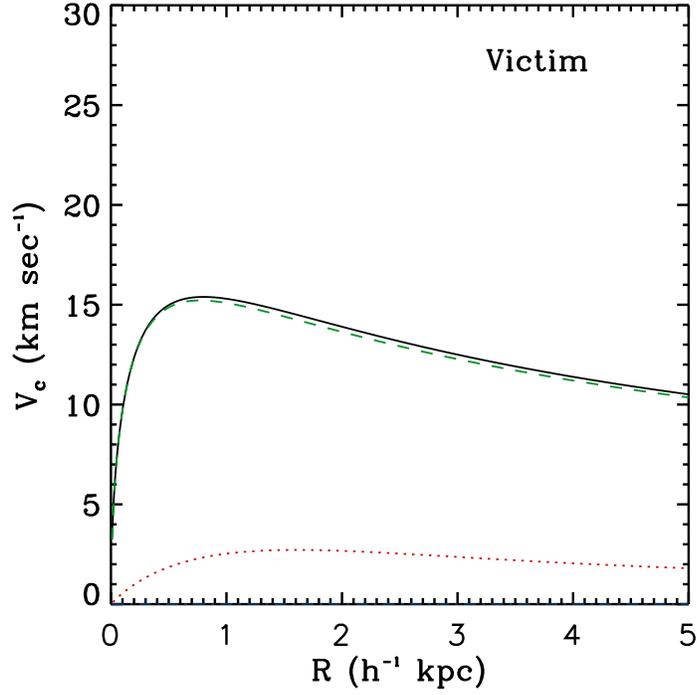}}
\vskip 0.2cm
\caption{Rotation curve for the small dwarf galaxy model 
with the following parameters:
  $v_{\rm{200}}=10$ km s$^{-1}$, c=25 (the dotted line displays 
the disk component, the dashed line the dark matter component and the 
solid line the total rotation curve, dark matter plus disk). The stellar mass disk is assumed to be 5\% of the 
total dark matter.  }  
\end{figure}

\begin{figure}
\vskip 8.5cm
{\includegraphics{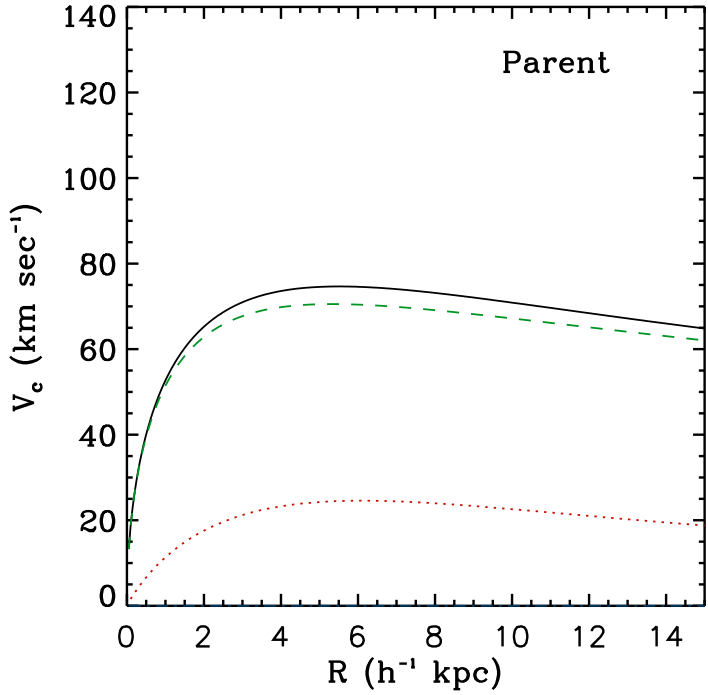}}
\vskip 0.2cm
\caption{Rotation curve for the modeled larger dwarf galaxy 
with the following
  parameters: $v_{200}=$50 km s$^{-1}$, c=20 (line types as above).}
\end{figure}

\begin{figure}
\vskip 8.5cm
{\includegraphics{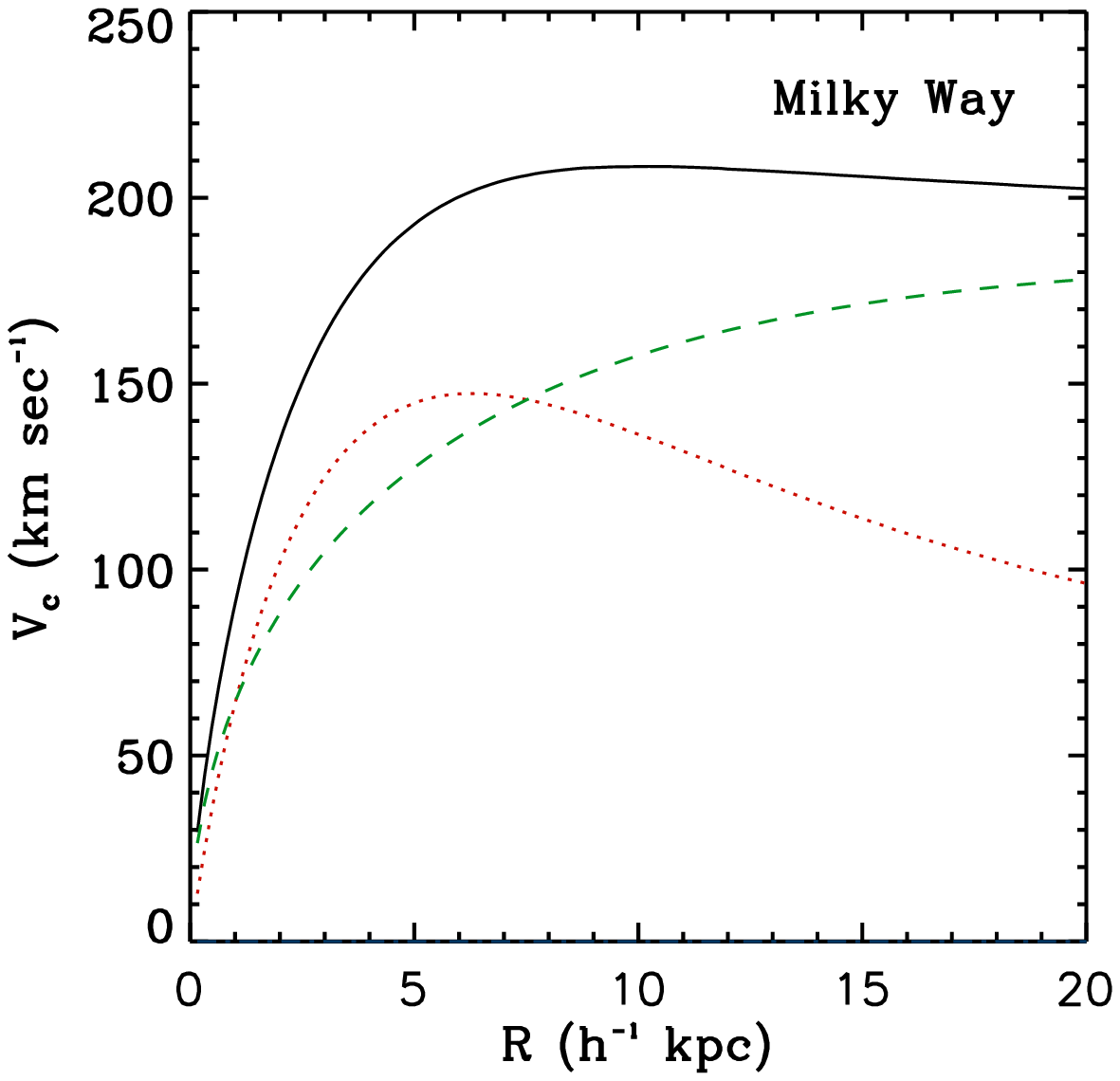}} 
\caption{Rotation curve for the  Milky Way model with the following
  parameters: $v_{200}=$160 km s$^{-1}$, c=9 (line types as above).}
\end{figure}

\section{Disks}

We model the 
disk component of the 
stars in each interacting galaxy with a
thin exponential surface density profile
of scale length $r_h$:\\ 
\begin{equation}
\Sigma_{*}(r)=\frac{M_{*}}{2\pi r_h^2} \rm{exp}(-r/r_h)
\end{equation}
so that the disk mass is $M_*=m_dM_{\rm{tot}}$, where $m_d$
is dimensionless and $M_{\rm{tot}}$ is the total mass of the galaxy.

The characteristic 
kinematic property of the small dwarf, $v/\sigma$, the ratio between
its centrifugal rotational velocity $v$ and its random motions $\sigma$,  
is measured at the effective
radius at each output time. The effective radius $R_e$ is defined as the radius
enclosing half the projected mass surface density profile,
$I(R_e)=I(R)/2$, with:
\begin{equation}
I(R)=\int_0^R \Sigma r dr.
\end{equation}

The circular velocity of each galaxy is given by:
\begin{equation}
V_c^2(r)=\frac{GM_{\rm{dm}}(<r)}{r}+\frac{2GM_*}{r_h}y^2 x
[I_0(y)K_0(y)-I_1(y)K_1(y)].
\end{equation}
Here, $G$ is the gravitational constant, $y=r/(2r_h)$,
and $I_n$ and $K_n$ are Bessel functions.

We specify the vertical mass distribution of the stars in the disk by
giving it the profile of an isothermal sheet with a radially constant 
scale height $z_0$. The 3D stellar density in the disk is hence given by:

\begin{equation}
\rho_*(r,z)=\frac{M_*}{4\pi z_0 r_h^2} \rm{sech}^2 \Big(\frac{z}{2z_0}\Big) exp\Big(-\frac{r}{r_h}\Big). 
\end{equation}

We treat $z_0$ as a free parameter that is set by
the vertical velocity dispersion of the stars in
the disk and
fix the velocity distribution of the stars such that this scale height is
self-consistently maintained in the fully 3D potential of the galaxy 
model\cite{SDH04}. We adopt $z_0 = 0.1$ $h^{-1}$ kpc.

Figures 5-7 display the rotation curves for the three galaxy models
we employed in our simulations: a small dwarf
galaxy, a larger dwarf, and a representation of the Milky Way.
The total mass of each galaxy is given in terms of a virial velocity $v_{200}$.
We set the total mass as $M_{\rm{tot}}=v_{200}^3/(10GH_0)$.
The small dwarf has a peak 
velocity of 15 km $s^{-1}$, implying a
total galaxy mass 
of 1.7x10$^{8} h^{-1}$M$_{\odot}$, ($v_{200}=10$ km $s^{-1}$). 
The larger dwarf galaxy has a peak velocity of
75 km s$^{-1}$ corresponding to a mass of ~5x10$^{10} h^{-1}$M$_{\odot}$, 
($v_{200}=50$ km $s^{-1}$), making it 
comparable to the Large Magellanic Cloud (LMC).  

The galaxy model for the Milky Way has
a mass of 9.5x10$^{11}h^{-1}$M$_{\odot}$
implying a virial velocity $v_{200}=160$ km s$^{-1}$.
The disk mass of each galaxy is specified by $m_d$=0.05.
We have set up models with low-mass disks (where the stellar disk is 
included in the dynamics of the simulations but its mass is negligible)
and found that the stars exhibit 
the same response to the gravitational resonance.

We have also set up models for the small
dwarf with a slowly rising rotation curve by 
assuming a shallow dark matter density profile in the inner parts. 
For this purpose we have introduced the density profile for the dark matter 
with the following form\cite{Deh}:
\begin{equation}
\rho = \frac{3 \rm{M}}{4 \pi} \frac{a}{(r+a)^4}.
\end{equation}
This profile  belongs to the same family of density profiles 
as the Hernquist model, but it has a core.
The isotropic distribution function for the energy is given by\cite{Deh}:
 \begin{equation}
 f(\epsilon)=\frac{3\rm{M}}{2\pi^3(G \rm{M} a)^{3/2}}
  \Big(\sqrt{2\epsilon}\frac{3-4\epsilon}{1-2\epsilon}-3 \rm{arcsinh}\sqrt{\frac{2\epsilon}{1-2\epsilon}}\Big)
\end{equation}

We found that this mass distribution is more susceptible to tidal tail formation in agreement
with previous works\cite{DMH}.
In this specific case, 90\% of the stars are preferentially removed by resonant
stripping, resulting in the formation of a dwarf galaxy with a luminous to dark matter
ratio a factor of two higher than our previous calculations (see Figure 8).

\begin{figure}
\vskip 15cm
{\includegraphics{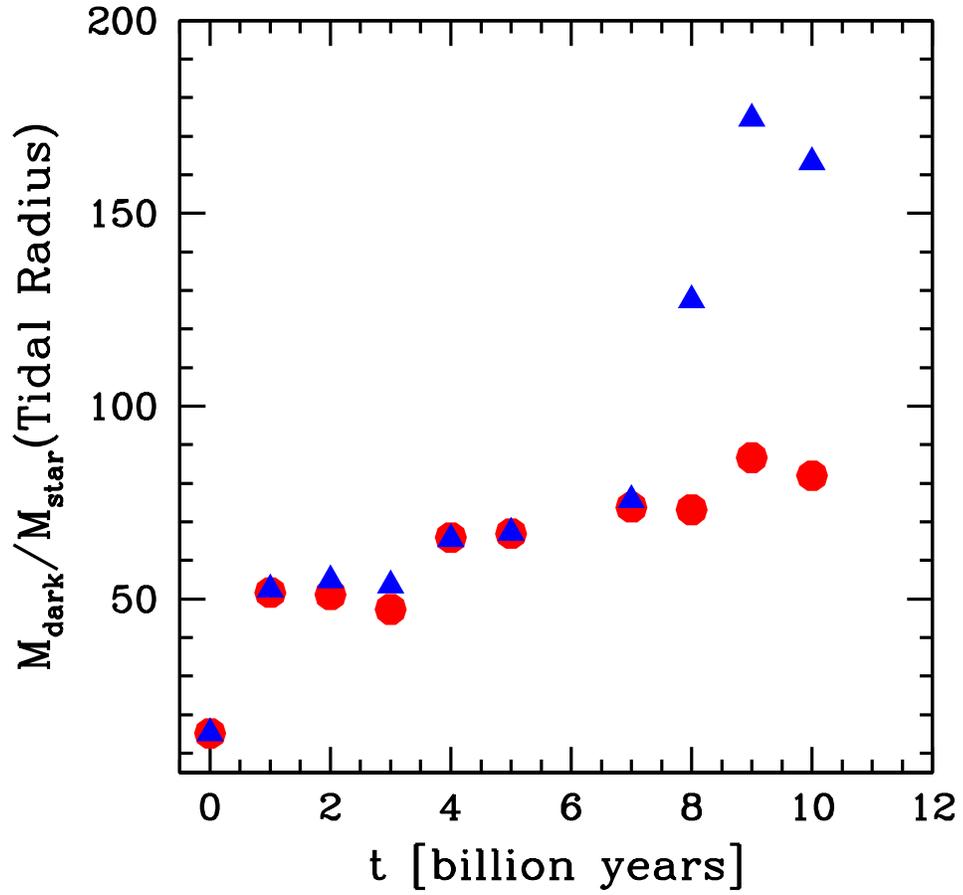}}
\vskip -6cm
\caption{The dark-to-luminous mass ratio (M$_{\rm{dm}}$/M$_{\rm{star}}$) computed at the tidal radius  
of the smaller dwarf under resonant stripping in a  prograde orbit assuming a 
shallow core for its dark matter density profile as  
function of time (blue triangles) as compared to a run where a cuspy 
central dark density
profile is assumed (filled red circles).}
\end{figure}

\section{Numerical Tests}

N-body simulations are subject to two-body relaxation effects.  These
may be particularly severe when several components with different mass
particles are present. Massive halo particles may heat up the disk
even in the absence of any external perturbation by colliding with
lighter stellar particles. If simulations have poor mass 
resolution, this numerical effect may heat a disk, thickening it with
time, and thereby artificially transforming it into a spheroid.  The models
presented here have a total of 10$^6$ particles to represent the dark
halos and 200,000 stellar particles in the disks.  As a
simple test, we have evolved the individual galaxies in isolation for
many billion years to see whether the discs remain thin or become thicker
because of numerical heating. Figure 9 shows that the disk of the
small dwarf in isolation (seen edge-on) is nearly static over the
course of 7 billion years, indicating that our models are not significantly
affected by numerical heating.

\begin{figure}
\vskip 9.2cm
{\includegraphics{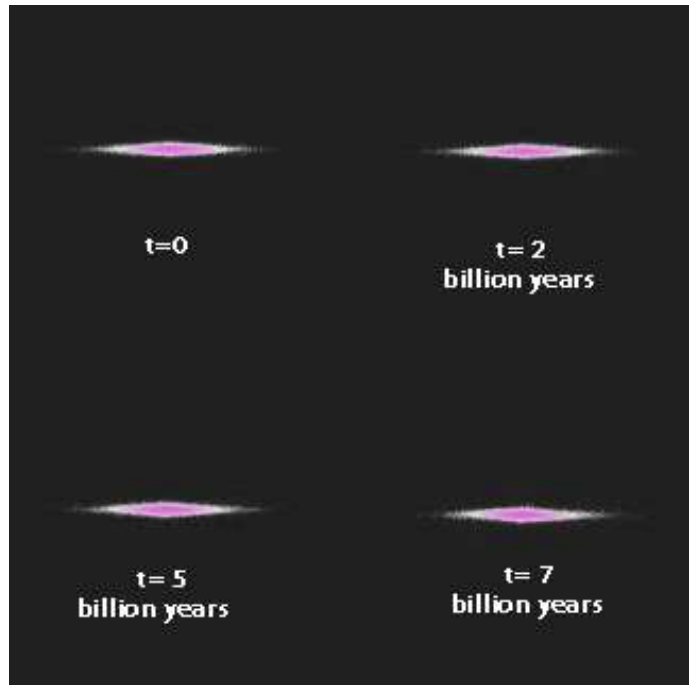}} 
\caption{Evolution in time of the disk of the small galaxy simulated in
  isolation without any external gravitational perturbation. The disk is
  displayed edge-on at the initial time and after 2, 5 and 7 billion years. 
The panels are 30 kpc on a side.  }
\end{figure}

\section{Resonant Response on Prograde and Retrograde Orbits}

In our simulations of a small dwarf interacting with a larger dwarf,
we place the galaxies on highly eccentric orbits (eccentricity e=0.8)
with a pericentric distance of $R_{\rm{peri}}$=5 $h^{-1}$ kpc and an apocentric
distance $R_{\rm{apo}}$=45 $h^{-1}$ kpc. These values are 
likely characteristic of
interactions between dwarf galaxies in the small group 
environments we are simulating. 
The disks for
prograde orbits are inclined with respect to the orbital plane by 30 and
60 degrees. 
We have also analyzed a coplanar configuration and the results are similar,
meaning that the choice of inclination is not critical to these results. 
For the simulations of a small dwarf interacting with
the Milky Way, the galaxies are placed on eccentric orbits (e=0.8)
with $R_{\rm{peri}}$=10 $h^{-1}$ kpc 
and $R_{\rm{apo}}$=100 $h^{-1}$kpc, and the disks
are inclined with respect to the orbital plane by the same angles as for
the dwarf interactions.  In contrast with heating mechanisms like tidal
shocking or stripping, the resonance that causes the
stripping depends less on the specific choice of the pericentric
distances but rather on the combination of the internal structure of the
small dwarf (the victim losing stars) and the orbital parameters (see
the main text).
\begin{figure}
\vskip 9.5cm
{\includegraphics{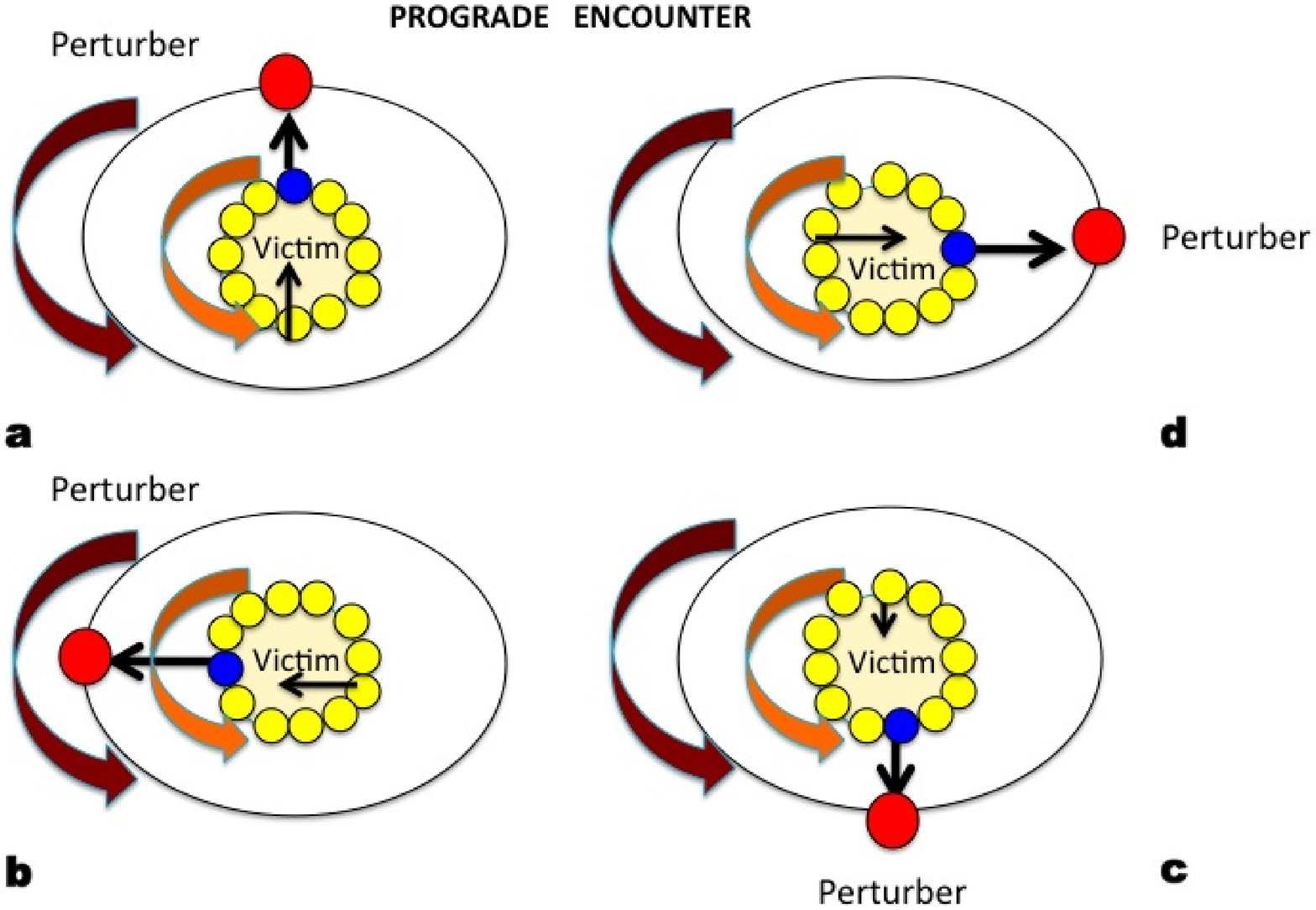}} 
\caption{ Resonant stripping response of the small dwarf galaxy 
(the victim) on prograde orbits under the influence of the 
gravitational perturbations from the 
larger dwarf galaxy (perturber). 
The sequence depicted above is counterclockwise.
For prograde encounters the star particles in the victim
(drawn as circles), which are rotating in the disk, resonate and 
are continuously pulled either inward or outward depending on
their initial positions in relation to the gravitational perturbation by
the larger galaxy.  As an example follow the path of the blue circle
in the victim.}
\end{figure}
\begin{figure}
\vskip 8.5cm
{\includegraphics{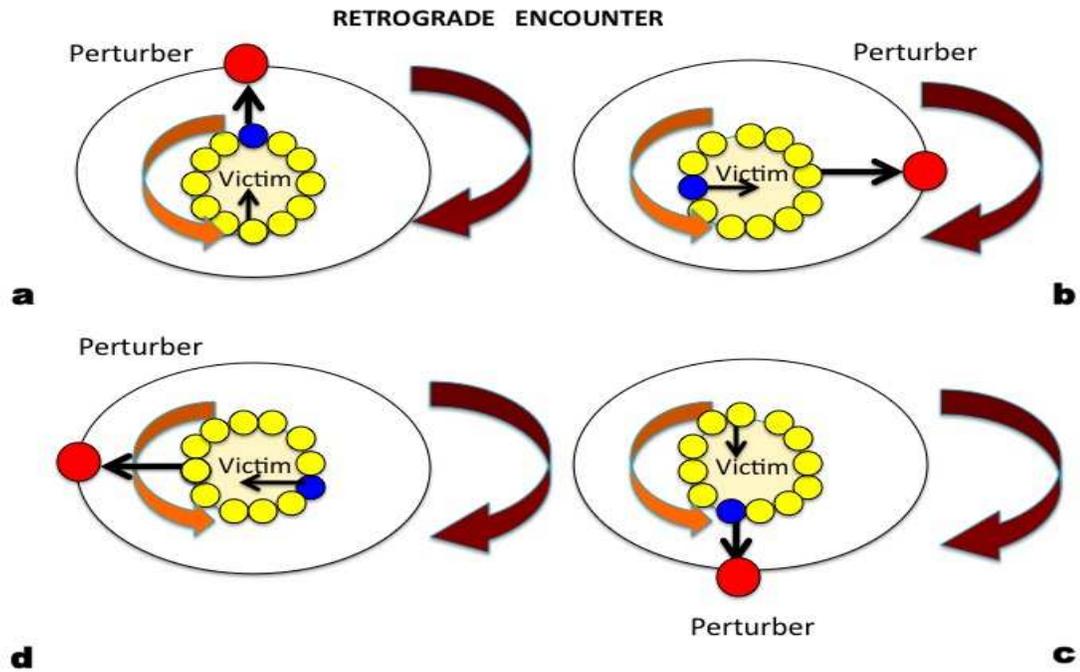}}  
\caption{Resonant stripping response of the small dwarf galaxy (the victim) 
on retrograde orbits under the influence of the gravitational 
perturbations from the 
larger dwarf galaxy (perturber). 
The sequence is clockwise.
For a retrograde collision, stars
in the disk of the victim are pulled alternately inward and outward (right panel) with little net
result.\cite{BT87}
As an example follow the path of the blue circle
in the victim.}
\end{figure}
\begin{figure}
\vskip 6.0cm
{\includegraphics{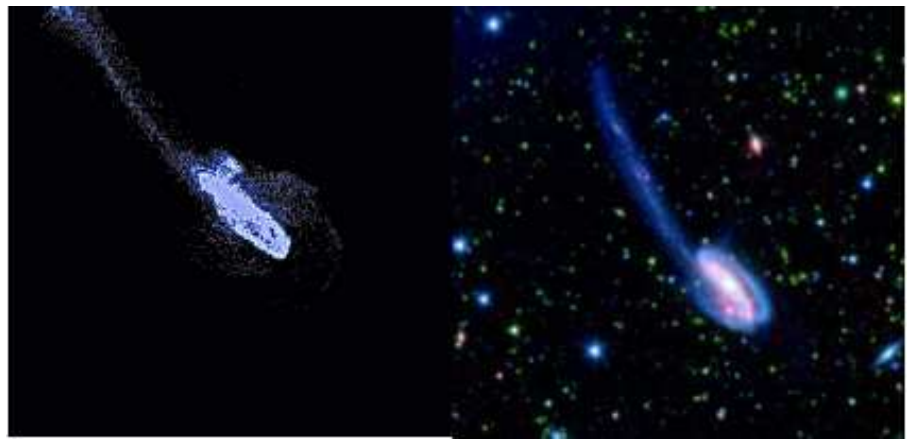}} 
\caption{Resonant stripping can produce
unusually extended tails (left panel) 
resembling the long tail observed in the
Tadpole galaxy (right panel).}
\end{figure}
\begin{figure}
\vskip 8.0cm
{\includegraphics{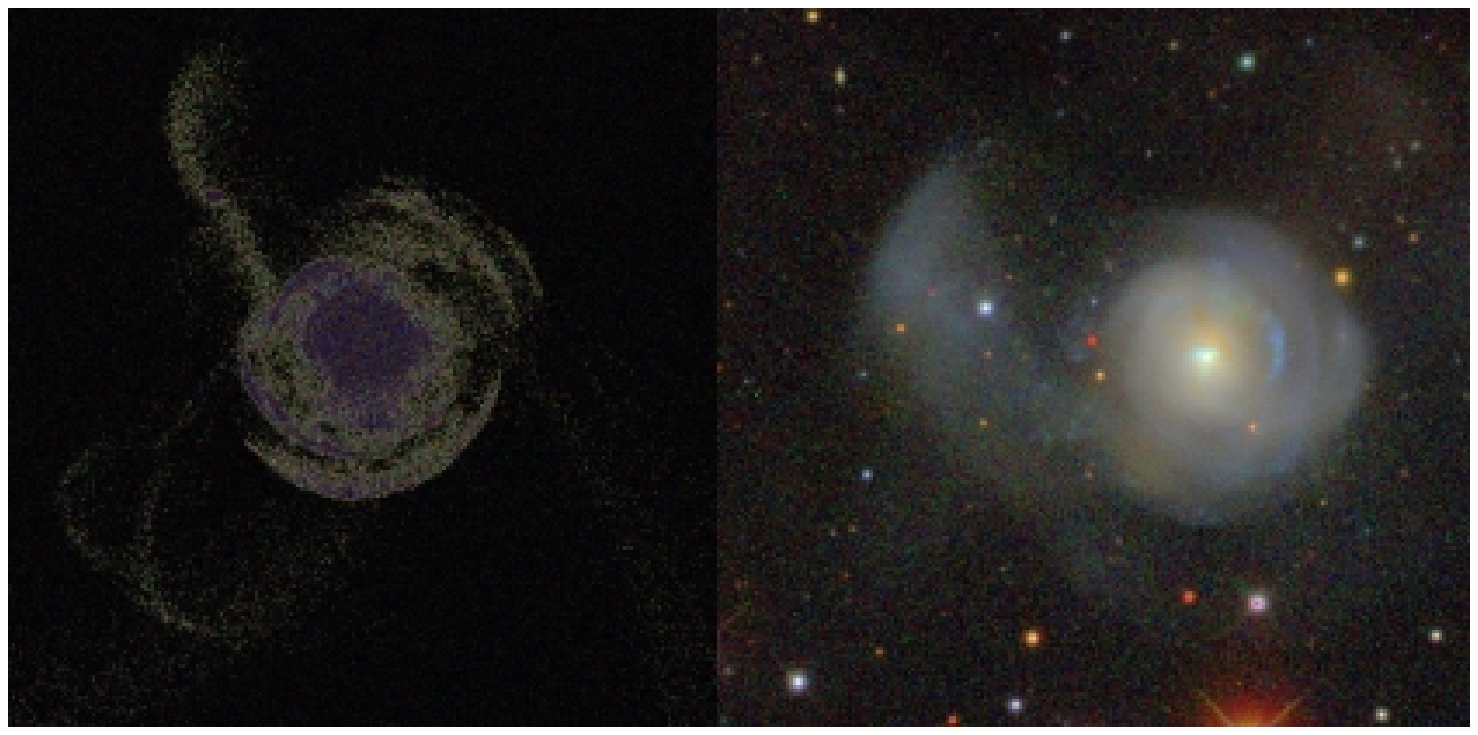}} 
\caption{Resonant stripping can produce shells in the 
larger dwarf galaxy owing to the capture of stars
stripped from the encounter with the  smaller dwarf. 
As an example we 
show an outcome of our simulation after a few billion years 
(left panel) which resembles the shells of stars
observed in NGC 2782 (right panel).}
\end{figure}
\noindent
Full details of the efficiency of this process during encounters between galaxies
with different internal structures, including different concentrations
and orbital configurations, need to be explored to generalize the applicability
of this process.

Resonant stripping is most effective for galaxies colliding 
on primarily
prograde orbits. In this case, as shown schematically in Figure 10-11,
stellar particles rotating in the disk of the small galaxy are in
resonance and are {\it continuously} pulled either inward or outward
relative to the center of the disk depending on their initial
positions in relation to the gravitational perturbation caused by
the larger galaxy (perturber) (Figure 10).  For a perfectly
retrograde collision, stars in the disk are pulled {\it alternately} inward
and outward (Figure 11) with little net result.\cite{BT87}  Because
the dark matter particles in the dwarf spheroidal candidate are on
random orbits, the net perturbation to the halo tends to average out,
greatly suppressing its response.

\section{Tails and Shells}

Resonant stripping can pull rotationally supported material 
like the stars in the dwarfs we
simulate, into extended tails and
bridges that resemble, e.g., the Tadpole galaxy (see Figure 12),
which is believed to have collided with a smaller companion.
Our simulations show that the curved tidal tail and shells in this
object can be explained by 
the process of resonant stripping, owing to an interaction with a system 100 times less massive. We argue that 
unusually long tails can be produced by the fact
that star particles rotating faster in the outer
part of the disk resonate strongly and are pulled into elongated shapes 
by the gravitational field of the larger galaxy. The distortion becomes so
large that half of the rotating material has been thrown far away from the
perturber forming a tidal tail behind 
the victim with the other half that has
been captured by  the perturber
forming arcs and shells of stars in 7 billion years resembling, e.g., 
the shells observed
in NGC 2782 (see Figure 13).

\end{document}